# Role of Data Mining in E-Payment systems

Sabyasachi Pattanaik, Partha Pratim Ghosh
FM University, Balasore

**Abstract**

*Data Mining deals extracting hidden knowledge, unexpected pattern and new rules from large database. Various customized data mining tools have been developed for domain specific applications such as Biomedicine, DNA analysis and telecommunication. Trends in data mining include further efforts towards the exploration of new application areas and methods for handling complex data types, algorithm scalability, constraint based data mining and visualization methods. In this paper we will present domain specific Secure Multiparty computation technique and applications. Data mining has matured as a field of basic and applied research in computer science in general. In this paper, we survey some of the recent approaches and architectures where data mining has been applied in the fields of e-payment systems. In this paper we limit our discussion to data mining in the context of e-payment systems. We also mention a few directions for further work in this domain, based on the survey.*

## 1. Introduction

E-payment has changed the face of most business functions in competitive enterprises. Internet technologies have faultlessly automated interface processes between customers and retailers, retailers and distributors, distributors and factories, and factories and their numerous suppliers. In general, e-commerce and e-business have enabled on-line Payment transactions. Also, generating large scale real-time data has never been easier. With data pertaining to various views of business transactions being readily available, it is only apposite to seek the services of data mining to make (business) sense out of these data sets. Data mining (DM) has as its dominant goal, the generation of non-obvious yet useful information for decision makers from very large databases. The various mechanisms of this generation include abstractions, aggregations, summarizations, and characterizations of data [1].

These forms, in turn, are the results of applying sophisticated modeling techniques from the diverse fields of statistics, artificial intelligence, and database management and computer graphics. The success of a DM exercise is driven to a very large extent by the following factors ([2]).

## 1.1 Availability of Data with Rich Descriptions

This means that unless the relations captured in the database are of high degree, extracting hidden patterns and relationships among the various attributes will not make any practical sense.

## 1.2 Availability of Large Volume of Data

This is mostly mandated for statistical significance of the rules to hold. Absences of say, at least a hundred thousand transactions will most likely reduce the usefulness of the rules generated from the transactional database.

## 1.3 Ease of Quantification of the Return on Investment (ROI) in Data Mining

Although the earlier two factors may be favorable, unless a strong business case can be easily made, investments in the next level DM efforts may not be possible. In other words, the utility of the DM exercise needs to be quantified vis-à-vis the domain of application.

## 1.4 Ease of Interfacing with Legacy System

It is commonplace to find large organizations run on several legacy systems that generate huge volumes of data. A DM exercise, which is usually preceded by other exercises like extract, transformation and loading (ETL), data filtering etc, should not add more overheads to system integration. It must now be noted that e-commerce data, being the result of on-line transactions, do satisfy all the above proper criteria for data mining.

We observe that once the back-end databases are properly designed to capture customer-buying behavior, and provided that default data take care of missing and non-existent data, the first issue of availability of data with rich descriptions is taken care of. Similarly, the reliability of data collected is also ensured because it is possible to increase the so-called no-touch-throughput in e-Payment transactions. Technologies like BizTalk and RosettaNet enhance the quality of data that is generated.

Improved web server availability results in faster transactions, thus increasing the revenue. Observe that increasing the number of transactions directly results in improved profits. Lastly, e-payment systems usually follow the MVC (Model-View-Controller) pattern with the business execution systems conforming to the model tier, the browser being the view tier and interfacing mechanisms like Java Servlets or Microsoft ASP forming the controller tier. Data mining mostly relies on the controller for generating the data to mine on. Thus integration issues also do not surface in this case. In summary, it is little surprise that e-payment transaction is the killer application for data mining ([3]).

## 2. A Review of Data-Mining Methods

The challenge in data mining is to disclose hidden relationships among various attributes of data and between several snapshots of data over a period of time. These hidden patterns have enormous potential in predictions and personalization in e-payment systems. Data mining has been pursued as a research topic by at least three communities: the statisticians, the artificial intelligence researchers, and the database engineers. We now present a brief overview of some of the features of each of these approaches.

### 2.1 Role of Statistics in Data Mining

Extracting causal information from data is often one of the principal goals of data mining and more generally of statistical inference. Data for decades; thus DM has actually existed from the time large-scale statistical modeling has been made possible.

Statisticians consider the causal relationship between the dependent variables and independent variables as proposed by the user (usually the domain expert), and try to capture the degree and nature of dependence between the variables. Modeling methods include simple linear regression, multiple regressions, and nonlinear regression. Such models are often parameter driven and are arrived at after solving attendant optimization models. For a more detailed overview of regression methods, the reader is referred to [3] & [4].

The regression methods may be considered analogous to the association rules in data mining. In the latter case, rule-mining algorithms propose the correlation of item sets in a database, across various attributes of the transactions. For instance, rules could be of the form if a customer visits Page A.html, 90% of the times she will also visit Page B.html. We assume here that the database (here, the web logs) has transactions recorded on a per-customer basis. Each record in the database indicates whether the customer visited a page during her entire session. Such rules can and need to be validated using the well-known statistical regression methods. Also, in some cases, the number of association rules may be very large. To draw meaningful rules that has real business value, it may be worthwhile to select the statistically most significant set of rules from the large pool of rules generated by a rule-mining algorithm. Data mining involves designing a search architecture requiring evaluation of hypotheses at the stages of the search, evaluation of the search output, and appropriate use of the results. Although Statistics may have little to offer in understanding search architectures, it has indeed a great deal to offer in evaluation of hypotheses in the above stages ([5]). While the statistical literature has a wealth of technical procedures and results to offer data mining, one has to take note of the following while using statistics to validate the rules generated using data mining.

- Prove that the estimation and search procedures used in data mining are consistent under conditions reasonably assumed to apply in applications.
- Use and reveal uncertainty and not hide it; some data-mining approaches ignore the causal relations due to lack of sufficient data. Such caveats can be unearthed using statistical methods.
- Calibrate the errors of search to take advantages of model averaging. This is relevant where predicting the future is important, as in data mining applied to forecasting a time series. Model averaging is beneficial where several models may be relevant to build a forecast.

### 2.2 The Role of AI in Data Mining

Artificial intelligence, on the other hand, has provided a number of useful methods for DM. Machine learning is a set of methods that enable a computer to learn relations from the given data sets. With minimal or no hypothesis from the user, learning algorithms do come up with meaningful relations and also explain them well. Some of the most popular learning systems include the neural networks and support vector machines. We briefly present the relevant issues below.

Neural networks are predominantly used to learn linear and nonlinear relationships between variables of interest. The architecture, in general, consists of a preceptor with input and output nodes with weighted edges connecting the two nodes. A neural network with two layers is thus a bi-partite acyclic graph. The preceptor, which is the learning machine, is 'trained' in order to arrive at an optimal 'weight vector'. The output is then expressed as a (weighted) linear combination of the inputs. Learning consists of solving an underlying optimization model which is solved using gradient descent based methods. It is worth noting here that the corresponding statistical methods available for estimating nonlinear relationships are based on the Maximum Likelihood Estimate problem. This problem is rather unwieldy since it requires the solution of highly nonlinear optimization problems; this result in tedious computations involved in solving algebraic equations. It is here that neural networks outperform their statistical counterparts, by resorting to the supervised learning methods based on gradient descent to solve such estimation problems. In other words, instead of explicitly solving equations to arrive at the maximum likelihood weights, neural networks 'learn' these weights via gradient descent based search methods. To learn more complex relationships including multi-variant nonlinear ones, it is not uncommon to have more layers than two. Such layers are called the hidden layers. The empiricism associated with neural networks is due to the non-availability of methods that would help fix the rate of convergence and the optimal number of layers. In the above learning process, if the outputs are Boolean, the problem is essentially a supervised learning mechanism to classify data sets. In such cases, often-times, a sigmoid function (a nonlinear transformation) is applied to obtain the relevant output. Apart from learning relationships as above, neural networks are also useful in clustering data sets. The most popular method available to cluster data sets is the K-means algorithm. Given an M-dimensional data set, the idea is to try and locate the minimal number of centroids around which the data set clusters itself. Thus the onus is to define an appropriate distance vector that helps partition the data sets into as minimally overlapping sub-sets as possible. The advantages of neural networks over the conventional statistical analysis methods are as follows ([6]).

- Neural networks are good at modeling nonlinear relationships and interaction while conventional statistical analysis in most cases assumes linear relationship between independent variables and dependent variables. Neural networks build their own models with the help of learning process whether the relationships among variables are linear or not.
- Neural networks perform well with missing or incomplete data. A single missing value in regression analysis leads to removal of the entire observation or removal of the associated variable from all observations in the data set being analyzed. However, neural networks update weights between input, output, and intermediate nodes, so that even incomplete data can contribute to learning and produce desired output results.

Neural networks do not require scale adjustment or statistical assumptions, such as normality or independent error terms. For a more detailed and comprehensive overview of neural computation and the underlying theories, the interested reader is referred to [7] & [8]

## 2.3 The Role of Database Research in Data Mining

Keeping in mind that data mining approaches rely heavily on the availability of high quality data sets, the database community has invented an array of relevant methods and mechanisms that
need to be used prior to any DM exercise. Extract, transform and load (ETL) applications are worthy of mention in this context. Given an enterprise system like an enterprise resource planning system (ERP), it is likely that the number of transactions that happen by the minute could run into hundreds, if not thousands. Data mining can certainly not be run on the transaction databases in their native state.
It requires be extracting at periodic intervals, transforming into a form usable for analysis & loading on to the servers and applications that work on the transformed data.

## 3. E-Payments and Data Mining
In this section, we survey articles that are very specific to DM implementations in e-payment systems. The salient applications of DM techniques are presented first. Later in this section, architecture and data collection issues are discussed.

### 3.1 DM in Customer Profiling
It may be observed that customers drive the revenues of any organization. Acquiring new customers, delighting and retaining existing customers, and predicting buyer behavior will improve the availability of products and services and hence the profits. Thus the end goal of any DM exercise in e-Payment is to improve processes that contribute to delivering value to the end customer.

### 3.2 DM in Recommendation Systems
Systems have also been developed to keep the customers automatically informed of important events of interest to them. The article by [9] discusses an intelligent framework called PENS that has the ability to not only notify customers of events, but also to predict events and event classes that are likely to be triggered by customers. The event notification system in PENS has the following components: Event manager, event channel manager, registries, and proxy manager. The event-prediction system is based on association rule-mining and clustering algorithms. The PENS system is used to actively help an e-commerce service provider to forecast the demand of product categories better. Data mining has also been applied in detecting how customers may respond to promotional offers made by a credit card company ([11]). Techniques including fuzzy computing and interval computing are used to generate if-then-else rules.

### 3.3 DM and Multimedia
Applications in virtual multimedia catalogs are highly interactive, as in e-malls selling Multimedia content based products. It is difficult in such situations to estimate resource demands required for presentation of catalog contents. [10] propose a method to predict presentation resource demands in interactive multimedia catalogs. The prediction is based on the results of mining the virtual mall action log file that contains information about previous user interests and browsing and buying behavior.

### 3.4 DM and Buyer Behavior in E-Payment Transactions
For a successful e-payment site, reducing user-perceived latency is the second most important quality after good site-navigation quality. The most successful approach towards reducing user perceived latency has been the extraction of path traversal patterns from past users access history to predict future user traversal behavior and to pre-fetch the required resources. The core of their approach involves extracting knowledge from integrated data of purchase and path traversal patterns of past users (obtainable from web server logs) to predict the purchase and traversal behavior of future users. In the context of web mining, clustering could be used to cluster similar click-streams to determine learning behaviors in the case of e-learning or general site access behaviors in ecommerce.

## 4. Data Collection and Software Architecture
### 4.1 Enabling Data Collection in E-Payment Systems

It may be observed that there are various ways of procuring data relevant to e-payments DM. Web server log files, web server plug-ins (instrumentation), TCP/IP packet sniffing, application server instrumentation are the primary means of collecting data. Other sources include transactions that the user performs, marketing programs (banner advertisements, emails etc), demographic (obtainable from site registrations and subscriptions), call centers and ERP systems.

### 4.2 An Architecture for DM

In a B2B e-commerce setting, it is very likely that vendors, customers and application service providers (ASP) (usually the middlemen) have varying DM requirements. Vendors would be interested in DM tailored for market basket analysis to know customer segments. On the other hand, end customers are keen to know updates on seasonal offerings and discounts all the while. The role of the ASP is then to be the common meeting ground for vendors and customers. [12] propose a distributed DM architecture that enables a DM to be conducted in such a naturally distributed environment. The proposed distributed data mining system is intended for the ASP to provide generic data mining services to its subscribers. In order to support the robust functioning of the system it possesses certain characteristics such as heterogeneity, costing infrastructure availability, presence of a generic optimization engine, security and extensibility. Heterogeneity implies that the system can mine data from heterogeneous and distributed locations. The proposed system is designed to support user requirements with respect to different distributed computing paradigms (including the client-server and mobile agent based models). The costing infrastructure refers to the system having a framework for estimating the costs of different tasks. This implies that a task that requires higher computational resources and/or faster response time should cost the users more on a relative scale of costs. Further, the system should be able to optimize the distributed data mining process to provide the users with the best response time possible (given the constraints of the mining environment and the expenses the user is willing to incur). The authors have indeed designed and implemented such a framework. Maintaining security implies that in some instances, the user might be mining highly sensitive data that should not leave the owner's site. In such cases, the authors provide the option to use the mobile-agent model where the mining algorithm and the relevant parameters are shipped to the data site and at the end of the process the mobile agent is destroyed on the site itself.

## 5. DM Applied to Retail Payment System

They share their experience in terms of lessons that they learnt. They classify the important issues in practical studies, into two categories: business-related and technology related. We now summarize their findings on the technical issues here.

- Collecting data at the right level of abstraction is very important. Web server logs were originally meant for debugging the server software. Hence they convey very little useful information on customer-related transactions. Approaches including sessioning the web logs may yield better results. A preferred alternative would be have the application server itself log the user related activities. This is certainly going to be richer in semantics compared to the state-less web logs, and is easier to maintain compared to state-full web logs.
- Designing user interface forms needs to consider the DM issues in mind. For instance, disabling default values on various important attributes like Gender, Marital status, Employment status, etc., will result in richer data collected for demographical analysis. The users should be made to enter these values, since it was found by [14] that several users left the default values untouched.
- Certain important implementation parameters in retail payment sites like the automatic time outs of user sessions due to perceived inactivity at the user end need to be based not purely on DM algorithms, but on the relative importance of the users to the organization. It should not turn out that large clients are made to lose their shopping carts due to the time outs that were fixed based on a DM of the application logs.
- Generating logs for several million transactions is a costly exercise. It may be wise to generate appropriate logs by conducting random sampling, as is done in statistical quality control. But such a sampling may not capture rare events, and in some cases like in advertisement referral based compensations, the data capture may be mandatory. Techniques thus need to be in place that can do this sampling in an intelligent fashion.
- Auditing of data procured for mining, from data warehouses, is mandatory. This is due to the fact that the data warehouse might have collated data from several disparate systems with a high chance of data being duplicated or lost during the ETL operations.
- Mining data at the right level of granularity is essential. Otherwise, the results from the DM exercise may not be correct.

## Conclusions and Future Work

In this paper, we have presented how web mining is applicable in improving the services provided by e-payment based enterprises. Statistics, AI and database methods were surveyed and their relevance to DM in general was discussed. Later, we also highlighted architectural and implementation issues. We now present some ways in which web mining can be extended for future work. With the growing interest in the notion of semantic web, an increasing number of sites use structured semantics and domain ontologies as part of the site design, creation, and content delivery. The primary challenge for the next-generation of personalization systems is to effectively integrate semantic knowledge from domain ontologies into the various parts of the process, including the data preparation, pattern discovery, and recommendation phases. Such a process must involve some or all of the following tasks and activities

- **Data transformations**: There are two sets of transformations that need to take place:
  (i) Data must be brought in from the operational system to build a data warehouse.
  (ii) Data may need to undergo transformations to answer a specific business question, a process that involves operations such as defining new columns, binning data, and aggregating it. While the first set of transformations needs to be modified infrequently (only when the site changes), the second set of
  transformations provides a significant challenge faced by many data mining
  tools today.
- **Scalability of data mining algorithms**: With a large amount of data, two scalability issues arise:
  (i) most data mining algorithms cannot process the amount of data gathered at
  web sites in reasonable time, especially because they scale nonlinearly.
  (ii) generated models are too complicated for humans to comprehend.